\documentclass[letterpaper, 10 pt, conference, xcolor={usenames}]{ieeeconf}  % Comment this line out if you need a4paper

\newcommand{\torque}{a}

\newcommand{\pendulumlength}{l}

\usepackage{lineno}
\usepackage{wrapfig}

\usepackage{hyperref}

\newcommand{\HP}[1]{{\color{NavyBlue}( #1 )}}

\IEEEoverridecommandlockouts                              % This command is only needed if 
\usepackage{graphics} % for pdf, bitmapped graphics files
\usepackage{epsfig} % for postscript graphics files
\usepackage{mathptmx} % assumes new font selection scheme installed
\usepackage{times} % assumes new font selection scheme installed
\usepackage{amsmath} % assumes amsmath package installed
\usepackage{amssymb}  % assumes amsmath package installed

\title{\LARGE \bf
{Controllability-Constrained Deep Network Models\\ for Enhanced Control of Dynamical Systems}
}

\author{Suruchi Sharma$^{1}$, Volodymyr Makarenko$^{1}$, Gautam Kumar$^{2}$, Stas Tiomkin$^{1,*}$\\{\small Charles W. Davidson College Of Engineering,
San Jose State University, CA, USA}\\{\hspace{0.7cm}\small\{suruchi.sharma, volodymyr.makarenko, gautam.kumar, stas.tiomkin\}@sjsu.edu}% <-this % stops a space
\thanks{This work was funded by Charles W. Davidson College Of Engineering (CoE),
San Jose State University (SJSU)}
\thanks{$^{1}$ Computer Engineering Department, CoE, SJSU
}
\thanks{$^{2}$ Chemical and Materials Engineering Department, CoE, SJSU
}
\thanks{$^*$ Corresponding Author}
}

\usepackage{amsmath}
\usepackage[dvipsnames]{xcolor}
\usepackage{graphicx} 
\usepackage[style=ieee,maxnames=99]{biblatex}
\usepackage{algorithm}
\usepackage{algpseudocode}
\usepackage{amsfonts}
\usepackage[normalem]{ulem}
\usepackage{siunitx}
\begin{document}

\makeatletter
\newenvironment{Pseudocode}[1][htb]{%
    \renewcommand{\ALG@name}{Pseudo-Code: Summary of the Proposed Method}% Update algorithm name
   \begin{algorithm}[#1]%
  }{\end{algorithm}}
\makeatother

\maketitle
\thispagestyle{empty}
\pagestyle{empty}

% \begin{figure*}
%   \includegraphics[width=\textwidth,height=7cm]{figures/schematic.png}
%   \caption{Workflow Diagram}
% \end{figure*}
%%%%%%%%%%%%%%%%%%%%%%%%%%%%%%%%%%%%%%%%%%%%%%%%%%%%%%%%%%%%%%%%%%%%%%%%%%%%%%%%
\begin{abstract}
% \color{blue}{}

% \TS{"enable to estimate" reads a bit strange I suggest "allow for the estimation of". thanks} 

Control of a dynamical system without the knowledge of dynamics is an important and challenging task. Modern machine learning approaches, such as deep neural networks (DNNs), allow for the estimation of a dynamics model from control inputs and corresponding state observation outputs. Such data-driven models are often  utilized for the derivation of model-based controllers. However, in general, there are no guarantees that a model represented by DNNs will be controllable according to the formal control-theoretical meaning of controllability, which is crucial for the design of effective controllers. This often precludes the use of DNN-estimated models in applications, where formal controllability guarantees are required. In this proof-of-the-concept work, we propose a control-theoretical method that explicitly enhances models estimated from data with controllability. That is achieved by augmenting the model estimation objective with a controllability constraint, which penalizes models with a low degree of controllability. As a result, the models estimated with the proposed controllability constraint allow for the derivation of more efficient controllers, they are interpretable by the control-theoretical quantities and have a lower long-term prediction error. The proposed method provides new insights on the connection between the DNN-based estimation of unknown dynamics and the control-theoretical guarantees of the solution properties. We demonstrate the superiority of the proposed method in two standard classical control systems with state observation given by low resolution high-dimensional images.

\end{abstract}

%%%%%%%%%%%%%%%%%%%%%%%%%%%%%%%%%%%%%%%%%%%%%%%%%%%%%%%%%%%%%%%%%%%%%%%%%%%%%%%%
\section{INTRODUCTION}

\begin{figure}[t!]
    \centering
    \includegraphics[scale=0.28,width=8.5cm]{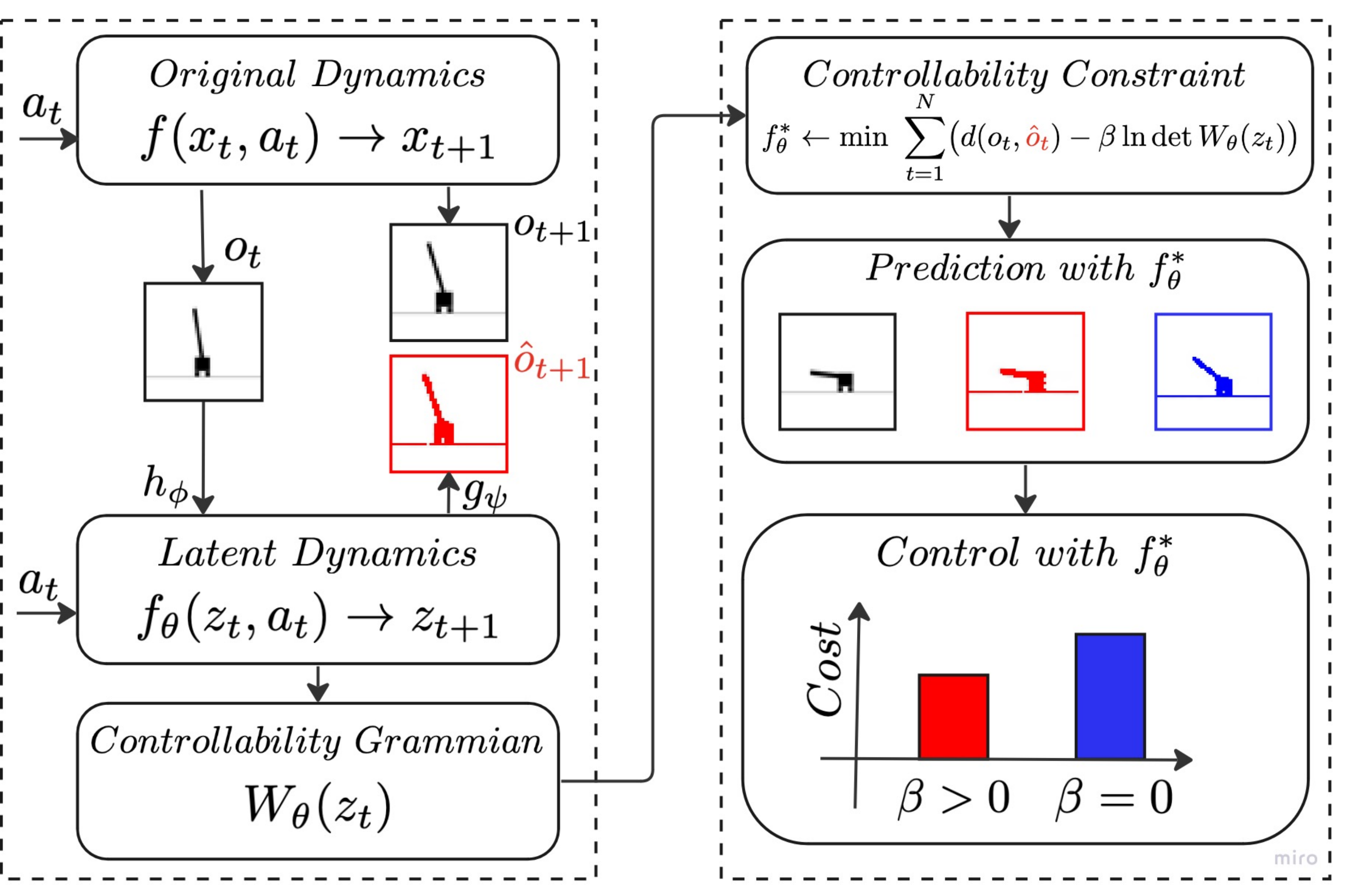}
    \caption{Method overview: the current and next time states, $x_t$ and $x_{t+1}$, in the original dynamics are rendered to the corresponding state-observations, given by low resolution images, $o_t$ and $o_{t+1}$, respectively. The former, $o_{t}$, is encoded to the latent state $z_t$ by the encoder, $h_{\phi}$. The next latent state, $z_{t+1}$, is predicted by the parametrized latent dynamics, $f_{\theta}$, and the corresponding predicted image, ${\color{red}\hat{o}_{t+1}}$, is generated by the decoder, $g_{\psi}$. The prediction error between the decoded and original state-observations, $\hat{o}_{t+1}$ and $o_{t+1}$, is minimized as in the standard VAE framework. We enhance the standard VAE framework with the controllability-constraint, which explicitly penalizes latent dynamics with low degree-of-controllability. The resulted controllability-constrained latent dynamics (in red) is superior to the standard baseline models (in blue) both in long-term prediction and control tasks. The former is visualized by a larger deviation of the standard model (in blue) from the original dynamics (in black), and the latter is depicted by a large control cost in the original dynamics driven by a controller derived with the standard model for $\beta=0$ in comparison to a controller derived with the controllability-constrained model for $\beta>0$.}\label{fig:scheme}
\end{figure}

{%\color{red}
% {\color{Green} reads good.  but does it speak to the control community? I think we can empahsize the control theoretical contribution right at the beginnging to set the appropriate tone}

% {\color{red} Agree. Gautam, feel free to edit in place in order to make it suitable for ACC. }

In recent years, the development of dynamical control models from high-dimensional data, such as images, using deep neural networks (DNNs) has garnered substantial interest within various fields, ranging from robotics to neuroscience \cite{cox2014neural,pierson2017deep}. Such data-driven models allow for the design of optimal control strategies for intricate dynamical control systems. However, these data-driven DNN-based models solely rely on the quality of input-output data acquired from a system. Often, this data lacks the desired control-theoretic properties, such as controllability, necessary for effective system control. Consequently, designing optimal controllers using these models can lead to sub-optimal performance. In this paper, we address this issue by explicitly incorporating a controllability metric into the development of latent models derived from high-dimensional imaging data. {By using these models within a model predictive control (MPC) framework, we demonstrate significant improvement in the MPC performance on two classical control systems: the inverted pendulum and a cart-pole system.}

% In recent years, the development of dynamical models from high-dimensional data (e.g., images) using deep neural networks (DNNs) approaches to develop optimal control strategies for complex systems has gained significant interest in robotics and neuroscience (Need references). These latent models solely rely on the quality of input-output data obtained from the system, which may lack the desired control-theoretic properties of the underlying input-output data, such as controllability, to control such systems effectively. As a result, the optimal controller design using these models can lead to poor performance. In this paper, we address this issue by explicitly incorporating a metric of controllability in developing latent models from high-dimensional imaging data.  

One of the widely adopted methods for capturing dynamical features in a low-dimensional latent space from high-dimensional data is Variational AutoEncoders (VAEs) \cite{kingma2013auto}. In essence, a VAE comprises an encoder model and a decoder model, both trained on the input data to minimize the reconstruction error between the encoded-decoded data and the input data \cite{kingma2013auto,watter2015embed}. Unlike standard autoencoders, which encode input to a single point, VAEs encode inputs as probability distributions across the latent space. VAEs have found applications in various domains, including {anomaly detection \cite{8819434}, image regeneration \cite{han2019variational}, data compression \cite{zhou2018variational,chamain2022end}}

Notably, VAEs have also emerged as valuable low-dimensional latent models for designing and controlling dynamical systems in recent research endeavors \cite{watter2015embed, banijamali2018robust}. {In this study, state information is represented  as images. Then Convolution Neural Networks are used for processing these  images and creating latent models. Convolution Neural Networks excel in handling image data, as demonstrated in  \cite{dosovitskiy2015learning, krizhevsky2012imagenet, hui2018research}. These  latent representations are used to acquire an understanding of the  dynamics of the environment, executing  control tasks similar to the approach in \cite{khalid1991neural,levine2019prediction, watter2015embed,banijamali2018robust}. Diving deeper into this concept reveals that this approach  not only helps in predicting short horizons but also  longer horizons  such as in \cite{hafner2019dream, hafner2019learning}.

None of the above approaches  explicitly include controllability constraints in their latent space. We aim to induce controllability constraints  in the latent space of the Variational AutoEncoder (VAE) model\cite{kingma2013auto}. The  Model Predictive Control techniques \cite{tassa2007receding, brunton2022data} are used to evaluate the performance of the models. %Since our approach relies on learned models, this work comes under the realm of model-based control, aligning with the methodologies in the studies  \cite{rybkin2021model, henaff2017model}.
} 
Since VAEs are trained using the input-output (control-observation) data from the system, the data used for training VAEs may limit the extent to which VAEs exhibit system controllability in response to exogenous inputs. Formally, the VAE's objective \cite{kingma2013auto} is to minimize the prediction error between the next state observation, generated by a dynamics model, given the current state observation and current control action. As we show in this work, this prediction error itself does not guarantee the controllability of a dynamics model, leading to the sub-optimal performance of controllers derived with such a model. 

Incorporating a controllability constraint, such as controllability Gramian \cite{brunton2022data}, during VAE training could significantly improve the intrinsic controllability of VAEs and lead to the development of efficient control strategies for complex systems. In this work, we conceptualize this idea by developing an approach to enhance the controllability of estimated models by imposing a controllability constraint to the standard VAE objective. {In particular, we augment the standard VAE objective for dynamics learning with the {\it degree of controllability} of the latent dynamics.} For our best knowledge, this is the first attempt to explicitly incorporate controllability constraints in designing latent models for dynamical systems estimated from data. 

Thus, this work establishes new connections between the state-of-the-art machine learning methods, (VAE), for the dynamics estimation from high-dimensional data (images/Lidar/etc) and the essential control-theoretical properties of dynamical control systems (controllability), leading to more reliable data-driven models and more efficient data-driven controllers.
}
The main contributions of this work are as follows:
\begin{itemize}
    \item Augmentation of the standard VAE objective for the estimation of unknown  dynamics from high-dimensional data with the controllability constraint.
    \item An efficient algorithm for the optimization of the controllability-constrained VAE objective.
    \item Demonstration of the effectiveness of the proposed method on the classical dynamical control systems on the tasks of long-term prediction and  model-predictive-control.
\end{itemize}
This paper is organized as follows. In Section \ref{sec:Problem Formulation}, we introduce the notations and necessary definitions for the main components such as state, state observation, action, original dynamics, $f$, and a dynamics model, $f_{\theta}$, (latent dynamics), degree of controllability and controllability Grammian, $W$, and feature extractor and image generator, (encoder, $h_{\phi}$ and decoder, $g_{\psi}$, respectively). Also, we overview the standard objective for learning a dynamics model from high-dimensional state observation in the formalism of VAE. In Section \ref{sec:method} we formulate a novel controllability-constrained objective for VAE, which comprises a prediction error term and a penalty term for a reduced degree of controllability of an estimated dynamics model. In Section \ref{sec:experiments} we demonstrate the advantage of the controllability-augmented dynamics models in comparison to the standard (baseline) model. Finally, in Section \ref{sec:conclusion} we conclude the current work and discuss future research directions and eventual applications. An overveiw of the method is provided at Fig. \ref{fig:scheme}.

\section{Problem Formulation}\label{sec:Problem Formulation}
\subsection{Original  and Latent Dynamics}
Consider a dynamical control system in discrete time, $f$:
\begin{align}
    f\;:\; x_t \times a_t \rightarrow x_{t+1}&&\text{(Original Dynamics)}\label{eq:real dynamics}
\end{align}
with the state and action at time $t$, denoted by $x_t\in\mathcal{X}$ and $a_t\in\mathcal{A}$, respectively. 

We assume the original dynamics function, $f$, is unknown, and the state, $x_t$ is partially observed through the state observation, $o_t\in\mathcal{O}$. The latter is given by a signal from a particular sensor, such as a camera, lidar, proprioceptive sensor, etc. The original dynamics can be examined by input-output triples, $\{o_t, a_t, o_{t+1}\}_{t=1}^N$, which are available or can be collected.

The goal of this work is to improve the quality of dynamics models estimated from data, $f_{\theta}$, {denoted by `latent dynamics' in the sequel, and the efficiency of controllers, $\pi$, derived with latent dynamics models.}

The latent dynamics is described by the following mapping: 
\begin{align}
    f_{\theta}\;:\; z_t \times a_t \rightarrow z_{t+1}\label{eq:latent dynamics}&&\text{(Latent Dynamics)}
\end{align}
where $z_t\in\mathcal{Z}$ is the latent state at time $t$, and $\theta\in{\Theta}$ is a model parameterizing dynamics in the model class, ${\Theta}$. 

A model, $\theta$, of latent dynamics, $f_{\theta}$, is estimated from data $\mathcal{D}=\{o_t, a_t, o_{t+1}\}_{t=1}^N$, by standard techniques \cite{watter2015embed, banijamali2018robust}, as overviewed below. In this work we assume that $\forall t\;:\;o_t$ is the state observation given by a low-resolution image, rendered from the original state, $x_t$.

% Latent dynamics, $f_{\theta}$, can be estimated from samples of control inputs, $a_t$ and corresponding state observations, $o_t\in\mathcal{O}$. The latter can be given by sensor recordings in various modalities such as camera images and/or LIDAR signals etc. 

\subsection{Estimation of Latent Dynamics}

Deep Neural Networks have been proven as a powerful model class, ${\Theta}$, for the representation of latent dynamics in this setting with an appropriate training objective \cite{hafner2019dream,hafner2019learning}:
\begin{align}
    \ell(\theta, \phi, \psi&) = \frac{1}{N}\sum_{t=1}^Nd\bigl({o}_{t+1}, \hat{o}_{t+1})\label{eq:dnn loss}\\
    &\mbox{with }\hat{o}_{t+1}=g_{\psi}\bigl(\underbrace{f_{\theta}(\overbrace{h_{\phi}({o}_{t})}^{z_t}, a_t)}_{z_{t+1}}\bigr),\nonumber
\end{align}
where $N$ is the number of training input-output pairs, $h_{\phi}$ and $g_{\psi}$ are trainable mappings to and from the latent space, respectively:
\begin{align}
h_{\phi}\;:\; & o_t\rightarrow z_t&&\text{(Encoder)}\\
g_{\psi}\;:\; & z_t\rightarrow o_t&&\text{(Decoder)}
\end{align}
and $d(\cdot, \cdot)$ is a given similar function between the original next state observation, $o_t\in\mathcal{O}$ and the reconstructed one, $\hat{o}_t\in\mathcal{O}$:
\begin{align}
    d\;:\; & o_t \times \hat{o}_t \rightarrow R^+.
\end{align} 
  
% {\color{red} Below paragraphs until Section C are not clear. Looks very vague.}

{
{%\color{Purple} PREVIOUS:

In our experiments, we used the commonly-used cross-entropy similarity between the original image, $o_t$, and the reconstructed image, $\hat{o}_{t}$, of the original dynamics. There exist various architectures and training methods for the derivation of $f_{\theta}$, $h_{\phi}$, and $g_{\psi}$, \cite{hafner2019dream, hafner2019learning}. 

Our method, (Section \ref{sec:method}) applies to arbitrary architectures and training methods, but in the experiments, we chose a particular latent dynamics model, 'Embed-To-Control', \cite{watter2015embed} due to its simplicity and transparency. 

An optimal solution to the problem in Eq.\eqref{eq:dnn loss} is derived by solving the following optimization problem   \cite{watter2015embed,banijamali2018robust}
\begin{align}
    \theta^*, \phi^*, \psi^* = \underset{\theta, \phi, \psi}{\mbox{ argmin }} \ell(\theta, \phi, \psi;\mathcal{D}),\label{eq:dnn objective}
\end{align}
where $\mathcal{D}=\{o_t, a_t, o_{t+1}\}_{t=1}^N$ is the training data.
There exist efficient techniques to solve the optimization problem in Eq.\eqref{eq:dnn objective} with high-dimensional observation and latent spaces \cite{tassa2007receding,li2004iterative}. In particular we used the 'ADAM' optimizer \cite{kingma2014adam}.

% Despite the success of DNNs for the representation of input-output mappings in various domains \cite{du2021learning}, 

A solution to Eq.\eqref{eq:dnn objective} does not guarantee, in general, controllability in estimated latent dynamics. In the next section,
we introduce a controllaiblty constraint to the optimization problem in Eq. \eqref{eq:dnn objective}, using {\it degree of controllability} as a metric \cite{brunton2022data}.

% we overview two controllability tests: a binary test - {\it a rank of controllability matrix}, and a more robust test - .
}

% induce into latent dynamics, $f_{\theta}$, the essential (for an efficient controller) features such as controllability. 

\subsection{Degree of Controllability}
In linear dynamics, $x_{t+1}=A_{n\times n} x_t+B_{n\times m} a_t$, controllability is given by a binary test, where the rank of the controllability matrix \cite{brunton2022data}, ${C}=[B, AB, A^2B, \dots, A^{n-1}B]$,  is equal to the state dimension, $n$, in a controllable system, and it is less than $n$ in an uncontrollable system. 
A more flexible controllability criterion is the degree of controllability, which is identified by the eigendecomposition of the controllability Gramian, $W_{n\times n}=C\cdot C^T$ \cite{brunton2022data}\footnote{The transpose is denoted by $T$, e.g., $M^T$ is the transpose of $M$.} More controllable directions, (eigenvectors of $W$), correspond to larger eigenvalues of $W$ \cite{brunton2022data}, and the  total degree of controllability can be identified by the controllability volume given by the determinant of the controllability Grammian. 

In non-linear dynamics, $x_{t+1}=f(x_t, a_t)$, the local degree of controllability can be estimated by the Grammian of locally-linearized dynamics around the state, $x$, which gives the local degree of controllability by the eigendecomposition of $W(x) = C(x)\cdot C(x)^T$. We use this observation in the development of the differentiable controllability constraint in the next section.

\section{Proposed Method for Controllability-Enhanced Latent Dynamics}\label{sec:method}
In this section we propose a novel method to explicitly induce controllability into latent dynamics, $f_{\theta}$, by augmenting the loss in Eq.\eqref{eq:dnn loss} with the controllability Grammian of locally-linearized latent dynamics around the latent state, $z$:
\begin{align}
    W_{\theta}(z) =& C_{\theta}(z)C_{\theta}(z)^T&&\hspace{-4.85cm}\text{(Controllability Grammian)}\label{eq:Grammian}\\ 
    C_{\theta}(z) =& [B_{\theta}(z), A_{\theta}(z)B_{\theta}(z), A^2_{\theta}(z)B_{\theta}(z),\dots \nonumber
    \\
 &\qquad\qquad\qquad\qquad\qquad\qquad\dots, A^{n-1}_{\theta}(z)B_{\theta}(z)]\nonumber
\end{align}
where $A_{\theta}(z) = \nabla_{z}f_{\theta}(z)$ and $B_{\theta}(z) =\nabla_{a}f_{\theta}(z)$ are the Jacobians of the latent dynamics, $f_{\theta}$, with regard to the latent state, $z$, and real action, $a$. And, $C_{\theta}(z)$ is the corresponding parametrized controllability matrix \cite{brunton2022data}.

In general, the pair $\bigl(A_{\theta}(z), B_{\theta}(z)\bigr)$ is not necessarily controllable, which may deteriorate the quality of control. In the next section, we augment the standard VAE-type loss for data-driven estimations of dynamics with the parametrized controllability Grammian, which explicitly penalized for a low degree of controllability.  
\subsection{Controllabilty-Enchanced Loss}
The parametrized Gramian, Eq. \eqref{eq:Grammian}, in the latent space, $W_{\theta}(z)$, allows us to formulate {\it the controllability-enhanced loss}, $\hat{\ell}$, for learning latent dynamics, $f_{\theta}$, as follows:
\begin{align}
    \hat{\ell}(\theta, \phi, \psi, \beta ;\mathcal{D}) = \ell(\theta, \phi, \psi ;\mathcal{D}) - \beta \ell_{\omega}(\theta;\mathcal{D})\label{eq:constraints}.
\end{align}
The second term in Eq.\eqref{eq:constraints} explicitly increases the total degree of controllability, (the eigenvalues of $W_{\theta}$), evaluated by the determinant of the controllability Grammian \cite{katayama2005subspace}
\begin{align}
    \ell_{\omega}(\theta;\mathcal{D}) \doteq \frac{1}{N}\sum_{t=1}^N \ln \det W_{\theta}(z_t)\label{eq:constraint}&&\text{(Constraint)}
\end{align}
{The interpretation of this constraint is the time-average controllability-volume of the linearized latent dynamics.} Importantly, this constraint is differentiable, which makes it suitable for gradient-based optimization \cite{kingma2014adam,ahmadianfar2020gradient}.
% The intuition of the {\it the controllability-enhanced loss} is as follows. The first term learns the mappings, between the observation space, $\mathcal{O}$, and the latent space, $\mathcal{Z}$, and the latent dynamics itself, $f_{\theta}$. The second term increases the degree of controllability of latent dynamics linearized around a nominal trajectory in the latent space.

The solution to Eq.\eqref{eq:constraints} can be derived by the same methods as in Eq. \eqref{eq:dnn objective}, with the essential difference that now $f_{\theta}$ has a desired degree of controllability set by $\beta$. 
\begin{align}
    \theta^*_{W}(\beta), \phi^*(\beta), \psi^*(\beta) = \underset{\theta, \phi, \psi}{\mbox{ argmin }} \hat{\ell}(\theta, \phi, \psi, \beta;\mathcal{D}).\label{eq:dnn ench solution}
\end{align}
Notably, in Eq. \eqref{eq:dnn ench solution}, the optimal parameters for latent dynamics, encoder, and decoder, are parametrized by $\beta$, which produces a family of solutions, as shown in Fig. \ref{fig:pendMPC}, and explained in the next sections.

The derivation of the optimal value of $\beta$ can be done by dual optimization (cf., Eq. (9) in \cite{peng2018variational} or in general \cite{nandwani2019primal}), which we defer to future work (cf., Section \ref{sec:conclusion}).

\subsection{Controllability-Prediction Trade-off}
\begin{figure}[t!]
\includegraphics[width=1.0\linewidth]{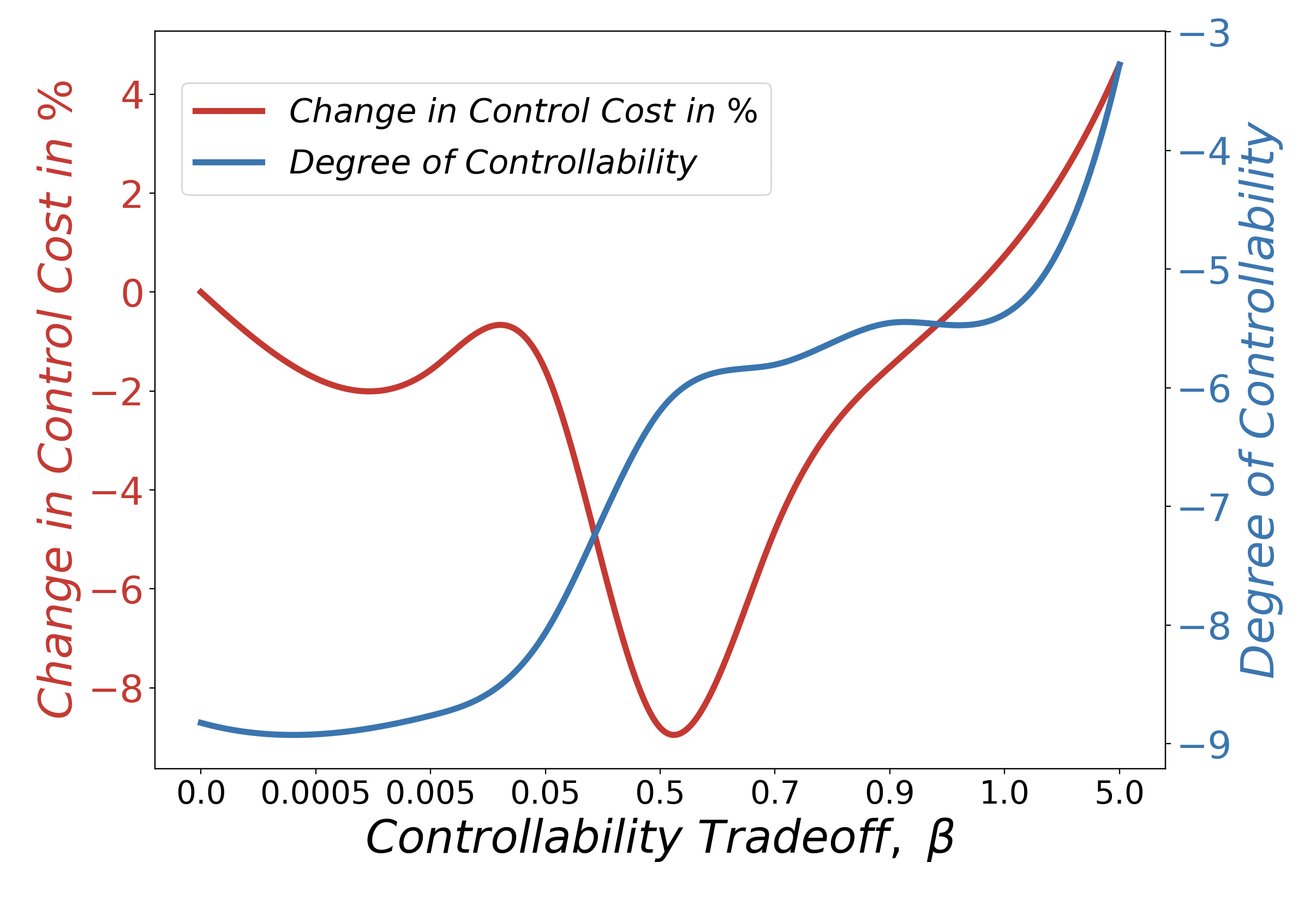}
\caption{Relative Change, in the MPC Cost (red) and Degree of Controllability (blue). The {\color{red}left y-axis} represents the Percentage change in the MPC cost for $\beta\in[0,\dots, 5.0]$. The MPC cost without controllability constraint is shown for $\beta=0$. There is $9\%$ decrease in the MPC cost for $\beta\approx 0.5$. The {\color{NavyBlue}right y-axis} shows the minimal eigenvalue of the controllability Grammian, $\ln\lambda_{\mbox{min}}(W(\beta))$, for the range of $\beta$ values, which reflects the degree-of-controllability,. There are $3$ orders of magnitude in the increase of the degree of controllability between $\beta=0$ and $\beta\approx 0.5$, which corresponds to the improvement in the efficiency of the MPC controller. As expected, for large values of $\beta$, the model predictability term in Eq. \eqref{eq:constraint} is less effective, resulting in a poor dynamical model for fitting the data.  \label{fig:pendMPC}}
  % \caption{\label{fig:pendMPC} Trade-off between the Controllability Degree and the Prediction Quality.  }
\end{figure}
% {\color {red} Figure 2 - need legend, x-axis tick mark font different, labels need to be clear. DONE}

The trade-off between the degree of controllability of latent dynamics, $z_{t+1}=f_{\theta}(z_t, a_t)$, and the quality of prediction of $o_{t+1}$ from $z_{t+1}=f_{\theta}(h_{\phi}(o_t), a_t)$ is controlled by a non-negative parameter, $\beta\ge 0$. This trade-off characterizes the relative importance between the first term in Eq. \eqref{eq:constraints} and its second term. The former
controls the overall prediction quality of the next image, $o_{t+1}$, from the current image, $o_t$, and the current action, $a_t$.The latter directly affects the degree of controllability of the parametrized latent dynamics, $f_{\theta}$. 

This way the optimal latent dynamics, $\theta^*_{W}(\beta)$, in Eq. \eqref{eq:dnn ench solution}, both possess a desired controllability degree and represents the mapping $\forall t\;:\;\{o_t\times a_t\rightarrow o_{t+1}\}$.

This trade-off produces a family of solutions, $f_{\theta^*(\beta)}$ parameterized by $\beta$, where the original solution is given by $\beta=0$. This additional degree of freedom in the optimization problem allows for better solutions with a particular property of interest in this work - controllable latent dynamics. We demonstrate this by numerical simulations of standard control benchmark systems in both tasks: planning with latent dynamics and deriving a more efficient controller for original dynamics. 

A typical trade-off, (family of solutions), is demonstrated in Fig. \eqref{fig:pendMPC}, where the solution without the controllability constraint in Eq. \eqref{eq:constraint} appears for $\beta=0$, and there is $9\%$ of the decrease in the Model-Predictive-Control cost for $\beta=0.5$. For larger $\beta$ values the MPC cost increases as expected because the first term in Eq. \eqref{eq:constraint} does not succeed in estimating the dynamics model. The MPC cost is estimated on control of the original dynamics, $x=f(x, a)$, while the optimal control actions are derived with the latent dynamics, $z=f_{\theta}(z, a)$. The full experiment settings are explained in the next section. 
% \subsection{Summary of the method}
\renewcommand{\thealgorithm}{}
\begin{Pseudocode}[t!]
  \caption{}\label{euclid}
  \begin{algorithmic}[1]
  \State { {Stage I: {Training controllability-enhanced models}} }
  \Require $\mathcal{D}=\{o_t, a_t, o_{t+1}\}_{t=1}^N$ - training data; $\beta$ - value of the trade-off parameter; $\theta$, $\phi$ and $\psi$ - initial DNN-parameters for the dynamics model, the encoder, and the decoder, respectively. 
  \Repeat
  \State {$\{\theta, \phi, \psi\} \leftarrow \{\theta, \phi, \psi\} - \nabla\hat{l}(\theta, \phi, \psi;\mathcal{D})$\Comment{Eq. \eqref{eq:constraints}}}
  \Until{convergence}
  \State \Return optimal ${f_{\theta^*}}, h_{\phi^*}, g_{\psi^*}$. 
  \State {\qquad {Stage II:{ Derivation of the optimal controller}} }
  \Require ${f_{\theta^*}}$, $h_{\phi^*}$, $x$, $x_g$, $H$\\
  // Encoding of original initial and target states
  \State $z= h_{\phi^*}\bigl(\mbox{RenderImage}(x)\bigr)$\Comment{initial latent state}
  \State $z_g= h_{\phi^*}\bigl(\mbox{RenderImage}(x_g)\bigr)$\Comment{target latent state}\\
  // Main Loop
  \Repeat
  % {\color{ForestGreen}
  \State {$a^* = \mbox{MPC}(\mbox{COST}, {f_{\theta^*}}, z, H)$\Comment{optimal action}}
  \State {$x= f(x,a^*)$\Comment{original dynamics}}
  \State {$z= {f_{\theta^*}(z,a^*)}$\Comment{latent dynamics}}
  % }
  \Until{$x=x_g$}
   \State \Return optimal control trajectory $\{a_1^*, a_2^*,\dots, a_k^*, \dots\}$
  \State \qquad\qquad {/* Utility functions: MPC and COST */}
  \Function{\mbox{MPC}}{\mbox{COST}, ${f_{\theta^*}}$, $z$, $H$}
  \State $\vec{a}\doteq\{a_1, a_2, \dots, a_{H-1}\} = \mbox{RandomActionSequence}$
   \State {$\vec{a}^*=\underset{\vec{a}}{\mbox{minimum}} \mbox{ COST}(\vec{a}, {f_{\theta^*}}, z, H)$}
    \State \Return $\vec{a}^*_1$\Comment{Return the first action}
  \EndFunction \mbox{ // MPC} \\
  //Inputs: latent state, action sequence, planning horizon
  \Function{\mbox{COST}}{$\{a_1, a_2, \dots, a_{H-1}\}, {f_{\theta^*}}, z, H$}
  \State $C = 0$
  \For{$k\in(1,\dots, H-1)$}
  \State {$e = z_g - z$\Comment{latent space error}}
  \State {$C = C + e^T\cdot e + a^T_k\cdot a_k$\Comment{accumulated cost}}
  \State {$z= {f_{\theta^*}}(z,a_)$\Comment{latent dynamics update}}
   \EndFor
    \State \Return $C$
  \EndFunction \mbox{ // COST}
  
   % \Function{Main}{}
   %   \State$a^*$=$MPC(TRANSIT(z_{t},a_{t}),z_{start},H)$
   %   \State $x_{t+1}$=$f(x_{t},a^*_{t})$
   %   \EndFunction
  
  \end{algorithmic}
\end{Pseudocode}

\subsection{Pseudo Code of the Method}
The proposed method is summarized by the {\bf Pseudo-Code} above, wherein the 'Stage I' a model, $f_{\theta}$, is estimated from data with the controllability-constrained objective (cf., Eq. \eqref{eq:constraints}), while in 'Stage II', this model is used for the derivation of the optimal controller to control an original dynamics, $f$, Eq. \eqref{eq:real dynamics}. In both cases, state observations are given by low-resolution images, rendered from the original state.

\section{Experiments}\label{sec:experiments}
In this section, we demonstrate the  method represented in  Section \ref{sec:method} on two classical control benchmarks, used for the evaluation of new algorithms for image-based estimation of dynamics and control. We evaluate the controllability-enhanced models, $f^*_{\theta}$, in two important tasks: long-term planning and model predictive control with the estimated model\footnote{All the experiments and results can be reproduced by our code repository:\url{https://github.com/suruchi1997/ControlledVAE.git}}. 

This section is organized as follows. Firstly, we explain the data collection and the evaluation metrics, then we provide the numerical simulations for 'Long-Term Planning' and 'Control' in two different dynamics: Inverted Pendulum and Cart Pole.

\subsubsection{Data Collection}

Data for training controllability-constrained models, $f_{\theta}$, is collected with the Open AI simulator for the classical control environments \cite{brockman2016openai, levine2019prediction,arroyo2021open}: 'Inverted Pendulum' and 'Cart Pole'. 

In data collection, the action $a_t$ is applied to the simulator of an environment, denoted by 'Step', which produces ('Renders') an image of the resulting state, $o_{t+1}$:
\begin{align}
    \forall t\in [0, \dots, N]\;:\:o_{t+1} = \mbox{RenderImage}(\mbox{Step}(a_t))\label{eq:render}.
\end{align}
This way the training data, $\mathcal{D}$, comprises of action and state-observation trajectories, $\mathcal{D}=\{a_t, o_t\}_{t=0}^N$.% The details of data collection are provided in Appendix \ref{sec1:firstappendix}. 

\subsubsection{Training} The training procedure is based on the standard stochastic gradient descent (SGD)  optimization \cite{kingma2014adam}, summarized by 'Stage I' in the {\bf Pseudo-Code}. 

In the experiments, the encoder, $h_{\phi}$, the latent dynamics, $f_{\theta}$, and the decoder, $g_{\psi}$, are implemented by a similar neural networks' architecture as in the baseline system \cite{watter2015embed}, which is a SOTA in the field. The architecture details are provided in Appendix  \ref{sec1:firstappendix} for completeness, and the exact implementations are accessible at the the link. Remarkably, our method requires only a small modification, (addition of the controllability constraint), to baseline systems, and can be eventually applied to arbitrary  systems. 

% Once data is collected, it is fed to the training environment, to generate learned models. The detailed description of the training process is shown in Algorithm 3 of the appendix section. Additionally, The Neural Network architectural summaries for each environment as well as all the generic and specific technical details are also provided in the appendix section. 

\subsubsection{Evaluation Metrics}
To examine the effectiveness of the controllability constraint Eq. \eqref{eq:constraint} on the quality of the latent dynamics, $f_{\theta}$, we used two metrics:

% \begin{itemize}
     {\it -Qualitative comparison} (cf., Fig. \ref{fig:pendImages} and Fig. \ref{fig:CartPoleImages}) between state observation trajectories generated by the ground truth dynamics in Eq. \eqref{eq:real dynamics} and by the trained latent dynamics, $f_{\theta}$, for different values of $\beta$.
    
     {\it -Quantitative comparison} (cf., Fig \ref{fig:pendMPC} and in the text below) between control cost by a controller derived with the latent dynamics, $f^*_{\theta}$, trained with the controllability-constraint, $\beta>0$ and without it $\beta=0$. 
    The MPC controller  \cite{845037, tassa2007receding}
    \begin{align}
        \pi: z_t\rightarrow  a_t\label{eq:controller}
    \end{align}is derived with the quadratic cost defined in the latent space, $\mathcal{Z}$, as summarized by 'Stage II' in the {\bf Pseudo-Code} (cf., lines 27-29). The optimal actions, $a^*_t=\pi(z_t)$, are applied to the original dynamics,
    \begin{align}
        x^*_{t+1} = f(x^*_t, a^*_t)
    \end{align}
    and {the quadratic cost for the optimal action sequence $\{a^*_1, a^*_2, \dots, a^*_k, \dots\}$ is calculated in the original space $\mathcal{X}$.}   
% \end{itemize}
   % and {\color{blue} the quadratic cost for  $J_{\mathcal{X}}[a^*_1, a^*_2, \dots, a^*_k, \dots]$ in the original space $\mathcal{X}$ is calculated.}   
These evaluation metrics address the two important properties of dynamics: 'Long-Term Planning' and 'Control', respectively.

\subsection{Inverted Pendulum}
\subsubsection{Original Dynamics}

 Swing up and stabilization of the inverted pendulum is one of the classic problems in control theory \cite{watter2015embed,muskinja2006swinging,levine2019prediction,bugeja2003non}. The goal of this experiment is to examine whether the controllability-constrained dynamics reflects better the intrinsic properties (long-term prediction and controllability) of the\begin{wrapfigure}[11]{l}{0.175\linewidth}\includegraphics[scale=0.1]{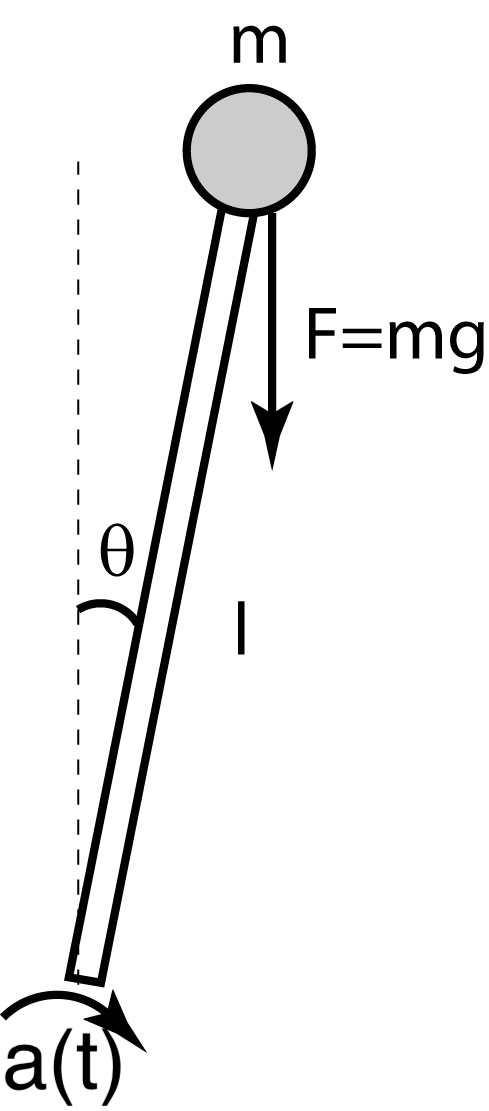}\end{wrapfigure}original dynamics. 
 
 In this experiment, we train the latent dynamics with low-resolution images, representing state observation, $o_t$, and real actions, $a_t$. This task is challenging because the model $f_{\theta}$ should learn a dynamics model from low-resolution  high-dimensional images. 
 
 The 'Inverted Pendulum' environment is schematically represented in the figure to the left. It comprised of a pivot point (the motor), where the limited control action (torque), $|a_t|<2  (\si{\newton\per\meter)}$, can be applied, and a pole with a point mass at the end.

 The state space of this system is two-dimensional, consisting of angle, $\theta_1=\theta (\si{\radian})$, and angular velocity, $\theta_2=\dot\theta (\si{\radian\per\second})$. The dynamics of the pendulum is given by:
\begin{align}
    \dot{\theta}_1(t) =& {\theta_2}(t) \nonumber\\
    \dot{\theta}_2(t) =& \frac{g}{\pendulumlength}\sin( \theta_1(t)) + \frac{\torque(t)}{m\pendulumlength^2}\label{eq:pendulum-equation}
\end{align}
where $g=9.8\si{\meter\per\second^{2}}$ is gravity, $m=1\si{\kilogram}$ is the mass of the point mass, and $l=1\si{\meter}$ is the length of the pole. 
% In the numerical simulations (data collection, planning, and control), we descretized Eq. \eqref{eq:pendulum-equation} in time with $dt=0.1\si{\second}$.

% In our case $m=1\si{\kilogram}$, $l=1\si{\meter}$, $g=9.8\si{\meter\per\second^{2}}$ and $dt=0.1\si{\second}$ that is rate of change in $\theta$ and $\dot\theta$ after a torque $a (\si{\newton\per\meter})$ is applied. 

\subsubsection{State Observation}

The dynamics in Eq. \eqref{eq:pendulum-equation} is\begin{wrapfigure}[7]{l}{0.35\linewidth}\vspace{-0.3cm}\frame{\includegraphics[scale=0.2]{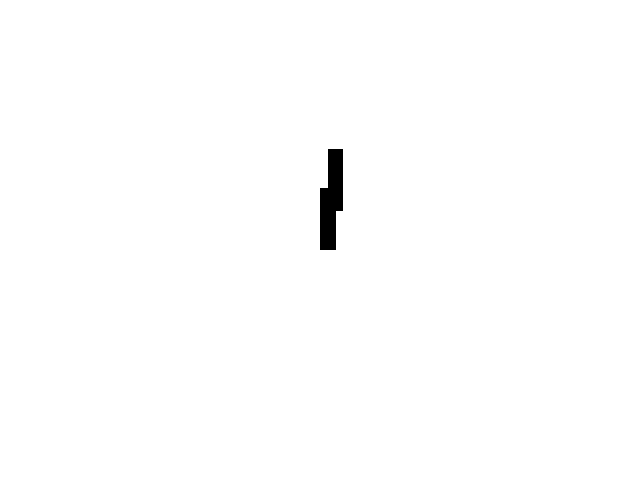}}\end{wrapfigure} simulated in discrete time with $dt=0.1\si{\second}$. The original state, $x_t$, in the time $t$ is rendered (cf., Eq. \eqref{eq:render}) to a corresponding low-resolution image with dimension $48\times 48$ pixels, which represents the state observation $o_t$, as shown to the left.

% \begin{figure}[h!]
%   \begin{minipage}{\linewidth}
%   \centering
%     \frame{\includegraphics[width=0.3\linewidth]{figures/pendulum/original/t8.png}}
%     \frame{\includegraphics[width=0.3\linewidth]{figures/pendulum/original/t9.png}}
%     \frame{\includegraphics[width=0.3\linewidth]{figures/pendulum/original/t10.png}}
%   \end{minipage} 
%   \caption{explain}\label{fig:state observation}
% \end{figure}

% For the inverted pendulum swing up task using trained latent environment{\color{red} incomplete sentence}. The state information is represented in form of images. The major drawbacks of the pendulum environment are firstly, the system is Non Markovian \cite{watter2015embed} and secondly, due to smaller size of input images (48$\times$48) there is a chance of discretization error while computing the angle of the pendulum \cite{watter2015embed}. Due to these reasons, capturing state information using a single image is really difficult. In order to solve this problem, two images are stacked together, forming  a single input image.This how the images for training data are generated. They are  utilized in training of both Controllability Constrained and Standard Models. 

% {\color{blue} Professor experiment settings and experiment results should be combined to one section because they talk about the results }

% \subsubsection{Experiment Results}

The qualitative comparison is shown in Fig. \ref{fig:pendImages}. The original dynamics (black), the controllability-constrained latent dynamics (red), and the baseline latent dynamics (blue) \cite{watter2015embed} start from the same initial state. However, in the three last rows, corresponding to the time steps $t_7$, $t_8$ and $t_9$, it is visible that the controllability-constrained dynamics better predict future states in comparison to the baseline dynamics. Additional details appear in the caption of Fig. \ref{fig:pendImages}. 

To validate the effectiveness of the proposed method for control, we derived an MPC controller in Eq. \eqref{eq:controller} with the controllability-constrained latent dynamics for $\beta\in[0, \dots 5]$. The change in the MPC cost for different values of $\beta$ is shown in Fig \ref{fig:pendMPC}. {Control cost in the original dynamics by the controller derived with baseline latent dynamics for $\beta=0$ is the reference}. The MPC cost is improved by $\sim 9\%$ by the controller derived with the controllability-constrained latent dynamics for $\beta\approx 0.5$. As expected, the MPC cost increases for larger values of $\beta$, because the controllability-constraint becomes dominant in the training objective Eq. \eqref{eq:constraints}, which deteriorates the prediction quality of the latent dynamics. This new {\it controllability-prediction trade-off} shows that estimated models can be enhanced by explicitly augmenting them with the controllability feature. This trade-off connects machine learning methods for sample-based model estimation with the fundamental concepts in optimal control theory.

 % as explained below. 
% Figure \ref{fig:pendImages} depicts how closely   predictions of the latent space replicate the next state predictions of the real environment when planned controls are applied. The black images represent the image trajectory for original dynamics. Images in red depict predictions from Controllability Constrained models (current work Eq.\eqref{eq:constraint}){\color{green}refer to a relevant equation}, and in blue depict predictions from Standard models (standard solution Eq.\eqref{eq:dnn loss}). For a trajectory that spans  10 steps starting  at  {\color{green}use siunit for degres}$45$${\si\degree}$ from the bottom,the images generated from latent space of Controllability Constrained and Standard models, are  similar at the  initial steps. However, in the final three steps,  it is clearly visible that Controllability Constrained models generate more accurate images as compared to Standard models

% the images, generated from latent space, are  similar to the initial steps. However, in the final three steps,  it is evident  that Controllability Constrained models generate more accurate images as compared to Standard models
% {\color{red} before it is mentioned as the same output}. 

% \subsubsection{Experiment Results}
\begin{figure}[t!]
\begin{center}
    \begin{tabular}{p{2cm}p{2cm}p{2cm}}
        \centering Original \\ Dynamics & \centering Constrained \\ Dynamics & \centering Unconstrained \\ Dynamics 
    \end{tabular}
\end{center}

  $t_0$\;\begin{minipage}{\linewidth}
    \frame{\includegraphics[width=0.3\linewidth]{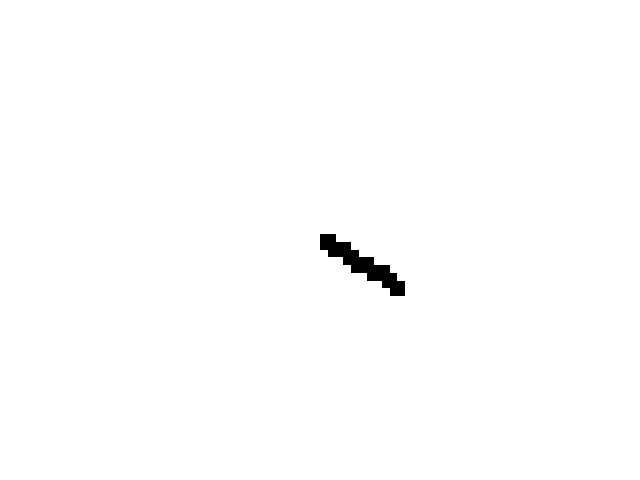}}
    \frame{\includegraphics[width=0.3\linewidth]{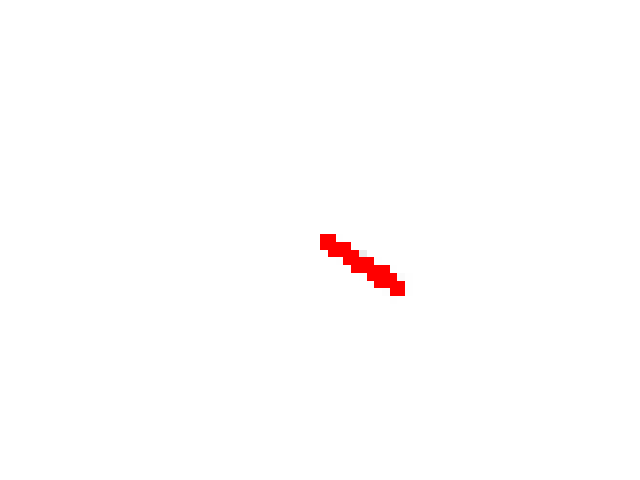}}
    \frame{\includegraphics[width=0.3\linewidth]{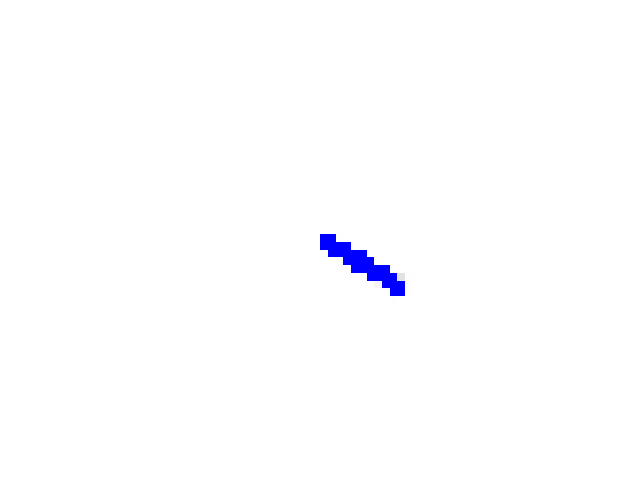}}
  \end{minipage} 
  \centering \\ 
  $t_2$\;\begin{minipage}{\linewidth}
    \frame{\includegraphics[width=0.3\linewidth]{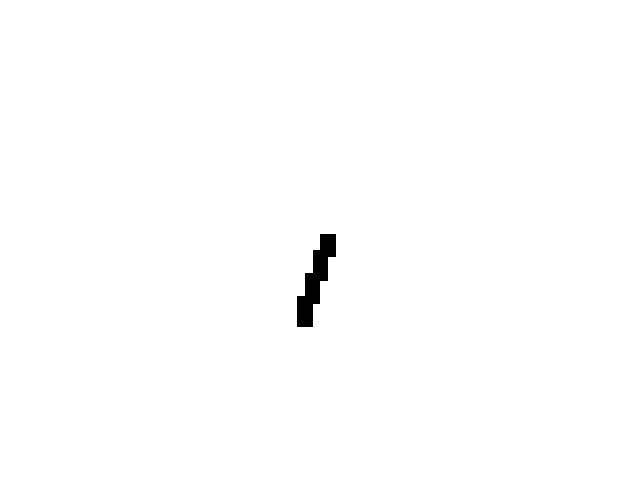}}
    \frame{\includegraphics[width=0.3\linewidth]{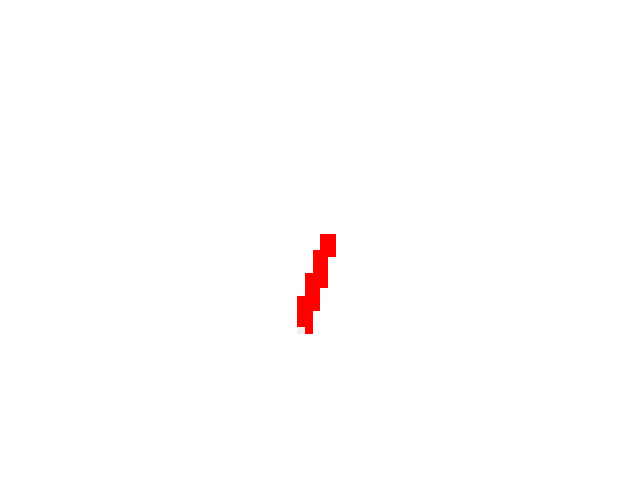}}
    \frame{\includegraphics[width=0.3\linewidth]{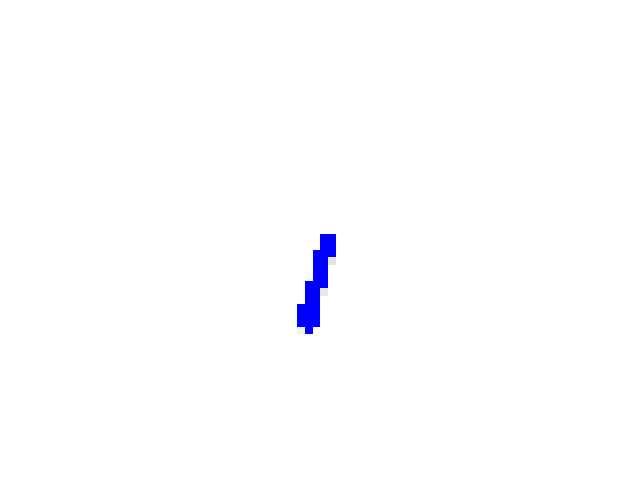}}
  \end{minipage} \\
  $t_7$\;\begin{minipage}{\linewidth}
    \frame{\includegraphics[width=0.3\linewidth]{figures/pendulum/original/t8.png}}
    \frame{\includegraphics[width=0.3\linewidth]{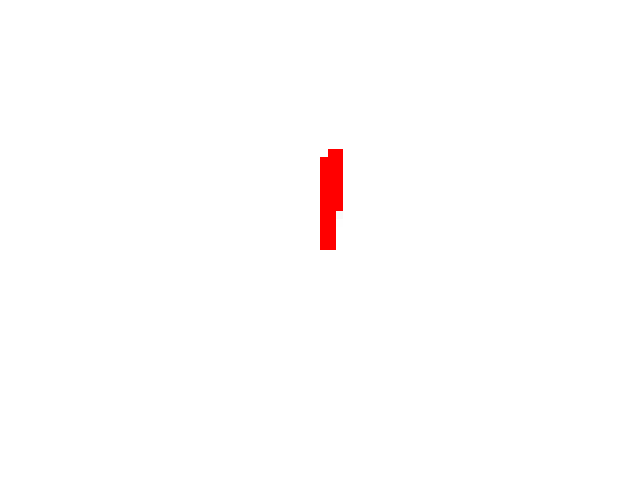}}
    \frame{\includegraphics[width=0.3\linewidth]{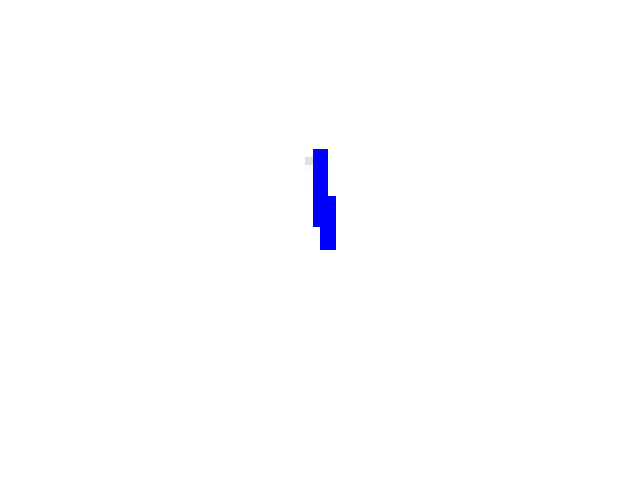}}
  \end{minipage}\\
  $t_8$\;\begin{minipage}{\linewidth}
    \frame{\includegraphics[width=0.3\linewidth]{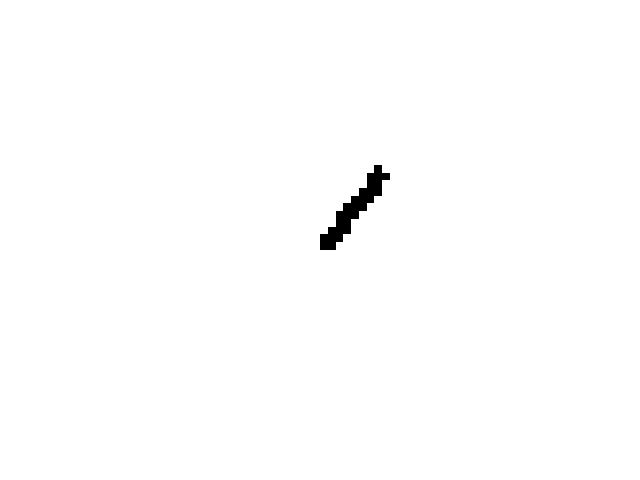}}
    \frame{\includegraphics[width=0.3\linewidth]{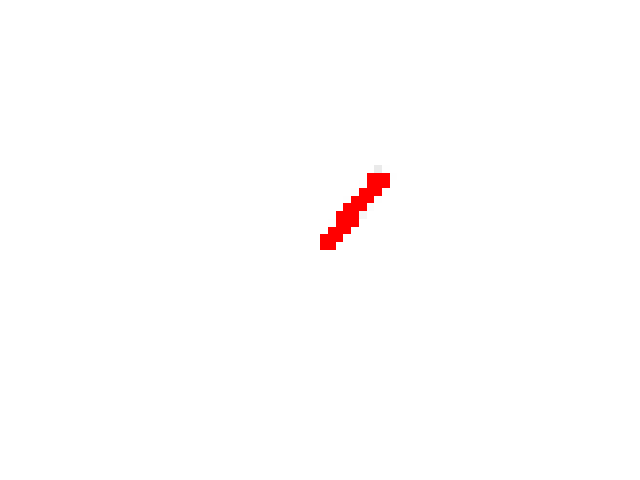}}
    \frame{\includegraphics[width=0.3\linewidth]{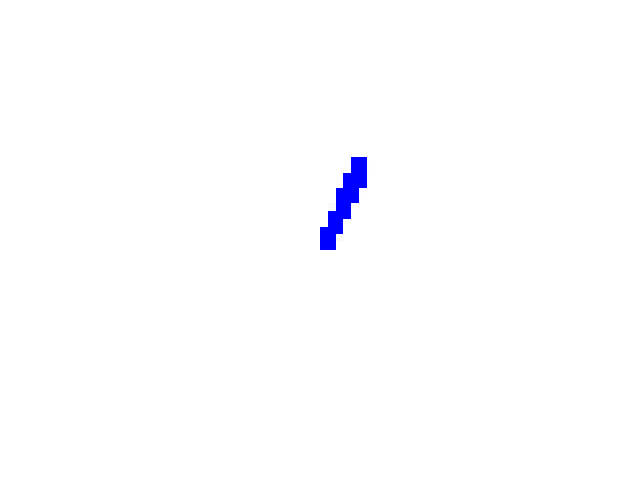}}
  \end{minipage}\\
$t_9$\;\begin{minipage}{\linewidth}
    \frame{\includegraphics[width=0.3\linewidth]{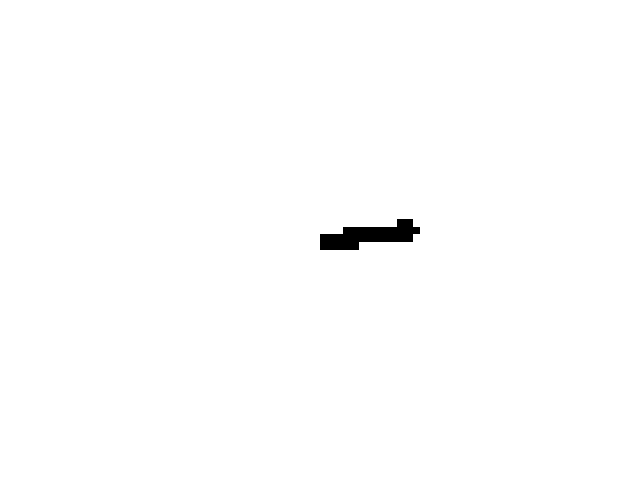}}
    \frame{\includegraphics[width=0.3\linewidth]{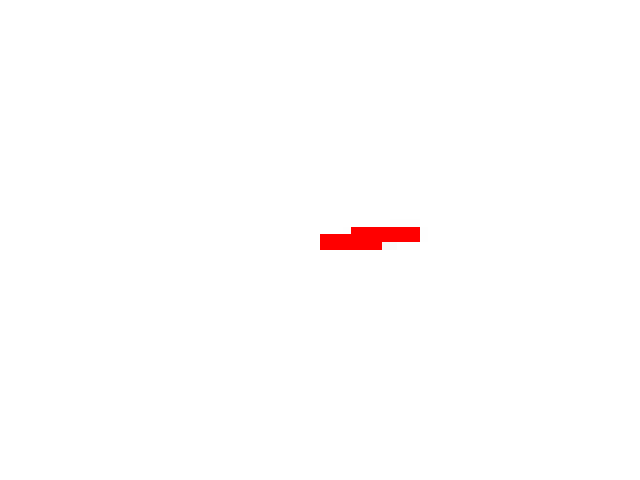}}
    \frame{\includegraphics[width=0.3\linewidth]{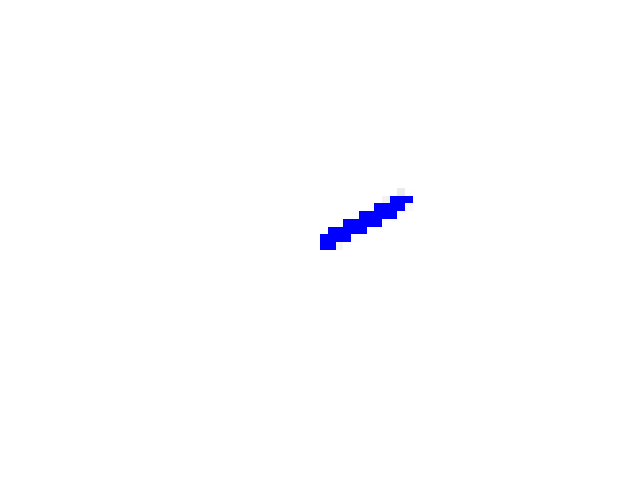}}
  \end{minipage} 
  % \textbf{$t_{13}$}\\
  \caption{\label{fig:pendImages}  Qualitative comparison between the long-term prediction by the latent dynamics, $f_{\theta}$, with the controllability-constraint (the 2nd column in red) and without it (the 3rd column blue). The first column represents the image trajectory by the original dynamics. The pivot point is at the center of the image. The pendulum starts at $\theta(t_0)=\frac{\pi}{4}\hspace{0.05cm}{\si{\radian}}$ from the bottom and $\dot{\theta}(t_0)=0\hspace{0.05cm}{\si{\radian\per\second}}$. The rows correspond to the time steps shown to the left. Visible changes appear after $7$ times steps. And, {as shown at the bottom row, after $9$ time steps, there is a significant deviation of the latent dynamics trajectory for $\beta=0$ (blue) from the original dynamics trajectory (black); while the latent dynamics trajectory for $\beta=0.5$ (red) closely follows the original dynamics. The deviation increases in further steps. }}
  % {\color{red} it would be good to put a label on top of each column - e.g., original dynamics, control without constraint etc., Same for Figure 4} DONE}
  % \caption{\label{fig:pendImages} Comparison between the prediction quality of latent dynamics with the controllability-constraint (current work, Eq. \eqref{eq:constraint}) and without controllability-constraint (standard solution, Eq.\eqref{eq:dnn loss}), represented in the second column in red and the third columns in blue, respectively. The first column represents the image trajectory by the original dynamics of the pendulum.}
\end{figure}
\subsection{Cart Pole}
To validate the effectiveness of the proposed method in a more complicated system, we conducted the same experiment with the 'Cart Pole' dynamics.
% {\color{green }  resolved- the same comments as in C. Pendulum. try to align these two section in their style/structure.}
\subsubsection{Original Dynamics} Balancing the pole over the cart is another classic problem in control theory.\begin{wrapfigure}[10]{l}{0.32\linewidth}\includegraphics[scale=0.1]{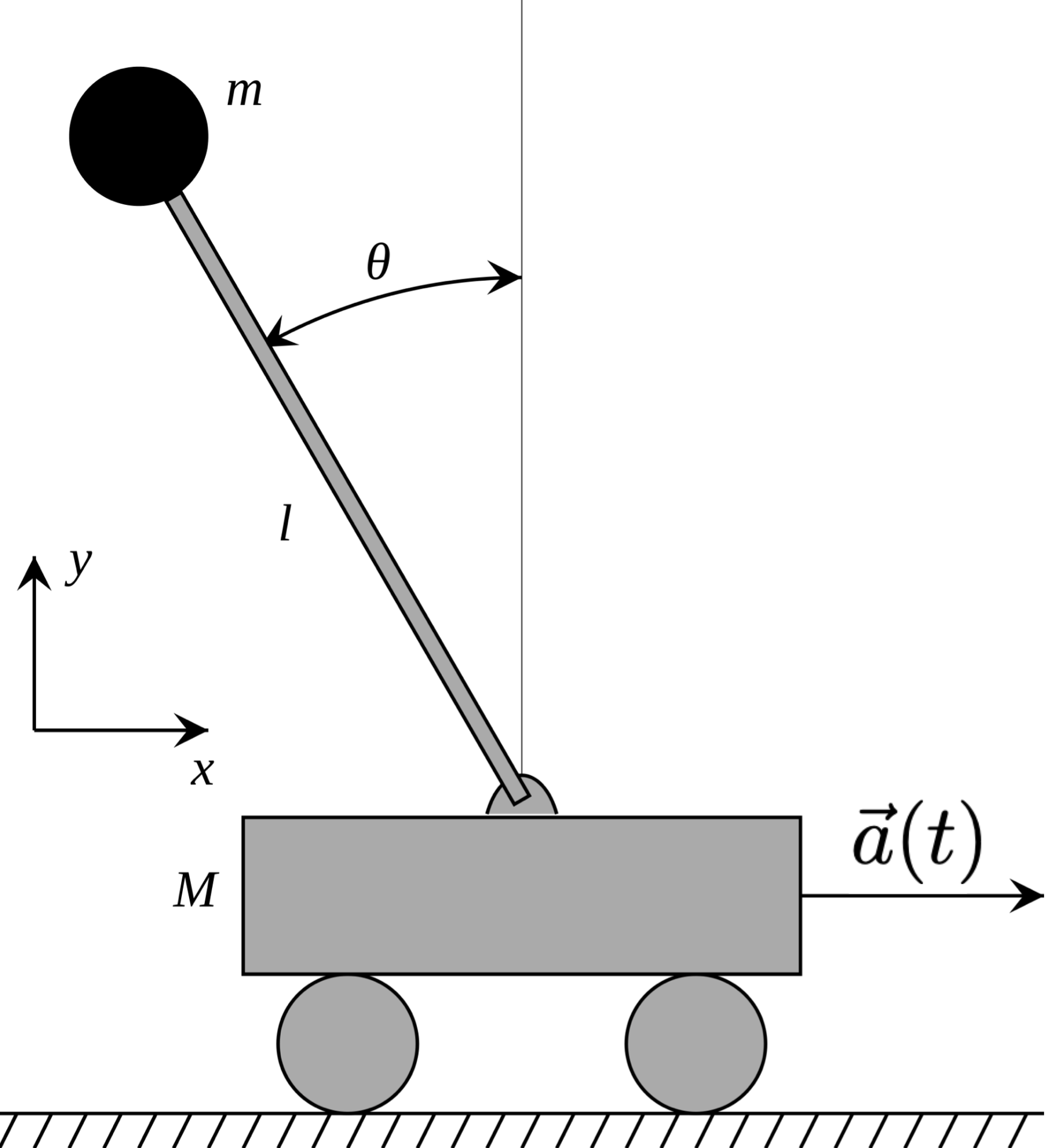}\end{wrapfigure}  
The cart pole system consists of a pole attached to a cart that moves over the frictionless track\cite{6313077}. The state space dimensionality is four. The components of state space include, cart position, ${x}$, cart linear velocity, $\dot{x}$, pole-angle, $\theta$, and pole angular velocity, $\dot\theta$.

The action space, $\vec{a}(t)\in\mathcal{A} $, is discrete, comprising of two force values +1${\si\newton}$ and -1${\si\newton}$. The original dynamics is given by
\begin{align}
\ddot{x}(t) =& \frac{m\sin\theta(t)(\ell\dot{\theta}^2(t)+g\cos \theta(t))+a(t)}{M + m\sin^2\theta(t)}\label{eq:cart-pole},\\
\ddot{\theta}(t) =& -a(t)\cos \theta(t) - m\ell\dot{\theta}^2(t) \cos\theta(t)\sin\theta(t)\nonumber\\
&\qquad\qquad\qquad\qquad\qquad\qquad\quad-(M+m)g\sin\theta(t)\nonumber,
\end{align}
where $x(t)$, $\theta(t)$, $m$, $M$, $\ell$, $g$, $|a(t)|\le 1$ are the $x$ coordinate of the center of mass of the cart, the angle of the pole, the pole mass, the cart mass, the pole length, the free-fall acceleration, and the force applied to the cart.
% \begin{align}
%     temp = (u + (m_{p}l)\dot\theta^{2}sin(\theta)) / (m_{p}+m_{c}) \label{eq:cpreal1}.
% \end{align}
% \begin{align}
%     \theta_{acc} = (g + sin(\theta) - cos(\theta)temp)  / (l+(4/3-m_{p}*cos(\theta^{2})/(m_p+m_c))) \label{eq:cpreal2}.
% \end{align}
% \begin{align}
%     x_{acc} = temp - m_{p}l\theta_{acc}cos(\theta)/(m_c+m_p)\label{eq:cpreal3}.
% \end{align}

% Using \ref{eq:cpreal1} \ref{eq:cpreal2} \ref{eq:cpreal3} we can compute the change in state components as

% \begin{align}
%     {x} = {x} + \dot{x}dt\label{eq:cpst1}.
% \end{align}
% \begin{align}
%     \dot{x} = \dot{x} + x_{acc}dt\label{eq:cpst2}.
% \end{align}
% \begin{align}
%     \theta = \theta + \dot{\theta}dt\label{eq:cpst3}.
% \end{align}
% \begin{align}
%     \dot{\theta} = \dot{\theta} + \theta_{acc} dt\label{eq:cpst4}.
% \end{align}

% where $g$=9.8 m/$s^{2}$ is gravity $m_{p}$ = 0.1 is mass of the pole , $m_{c}$ = 1.0 is the mass of the cart, l=0.5 is the length of the pole

\subsubsection{State Observation}
% \begin{figure}[h!]
%   \begin{minipage}{\linewidth}
%   \centering
%     \frame{\includegraphics[width=0.3\linewidth]{figures/cartpole/original/t1.png}}
%     \frame{\includegraphics[width=0.3\linewidth]{figures/cartpole/original/t3.png}}
%     \frame{\includegraphics[width=0.3\linewidth]{figures/cartpole/original/t8.png}}
%   \end{minipage} 
%   \caption{State observation in Cart Pole experiment. Left to right: observations at three consequent time steps, $o_{t_1}, o_{t_2}$ and $o_{t_3}$. The observation are $48\time 48$ pixel images, rendered from three corresponding states in equation-of-motion in Eq.\eqref{eq:cart-pole}}\label{fig:state observation cartpole}
% \end{figure}
In the same procedure as in the Inverted Pendulum, the dynamics in Eq. \eqref{eq:cart-pole} is\begin{wrapfigure}[8]{l}{0.35\linewidth}\frame{\includegraphics[scale=0.2]{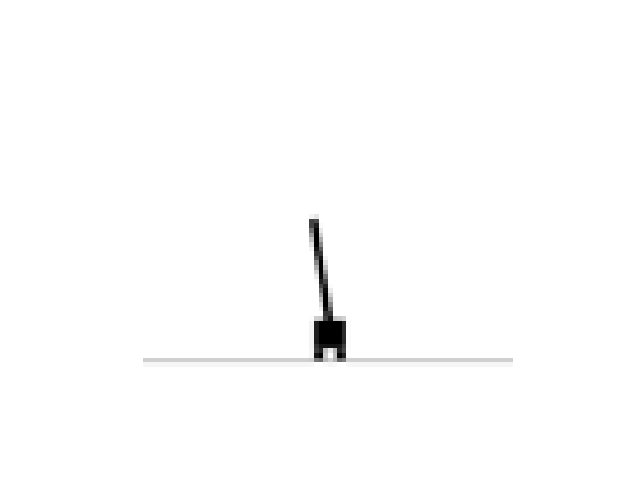}}\end{wrapfigure}simulated in discrete time, and the state $x_t$ at the time, $t$, is rendered to a low-resolution image with dimension $80 \times 80$, shown to the left.

% \subsubsection{Experiment Setting}
% For balancing the pole over the cart using trained latent environment. The state is rendered in the form of images. Due to discretization errors, capturing full-state information using a single image is difficult. In order to solve this problem, two images are stacked together, forming  a single input image.% This is how the images for training data are generated. They are  utilized in the training of both Controllability Constrained and Standard Models. 

Similarly to the 'Inverted Pendulum' environment, the performance is evaluated on both long-term prediction and control tasks.
Fig. \ref{fig:CartPoleImages} shows the qualitative comparison between the future state predictions by the latent dynamics and the original dynamics. The black images, in the 1st column, represent the image trajectory by the original dynamics. The red images, in the 2nd column, show the predictions by the controllability-constrained models, and the blue images, in the 3rd column, show the predictions by the baseline  models. All three dynamics start from the same initial conditions, the 1st row. The images generated by the controllability-constrained model (red) and the baseline model (blue), are similar at the first two steps. In the further steps, it is visible that controllability-constrained models accurately follow the original dynamics, while the baseline models (trained without the controllability constraint) deviate from the original dynamics. This deviation increases in further steps.

% Fig \ref{fig:CartPoleImages}  depicts the same result. We can see the black images represent the image trajectory for original dynamics. It is a 10-step trajectory starting in the upright position. Images in red depict predictions from Controllability Constrained models and in blue depict predictions from Standard models. The images, generated from latent space, are  similar at the initial steps. But  in the end, it is visible that Controllability Constrained models generate more accurate images as compared to Standard models. 

% \subsubsection{Experiment Results}

The control cost in the original dynamics driven by the controller derived with the controllability-constrained latent dynamics for $\beta=0.005$ is lower than that derived with the baseline latent model for $\beta=0.0$.

% the quadratic cost in the original dynamics 

% The real planning cost improves. We see a curve similar to the pendulum environment. The best range of $\beta$ values
% to generate controllability-constrained latent space varies
% between (0.002, 0.007) with minima close to 0.005.

% {\color{blue}The improvements are also clearly visible in trajectory calculations.  Fig \ref{fig:CartPoleImages} illustrates  how well images generated by Controllability Constrained models match the estimations from the real dynamics compared to Standard models, that too for an extended time horizon.}

% {\color{green}Suruchi: please explain here why Fig 4 is super "cool" and superior. Focus the readers on the advantage. and how to read the plot.}

% The cart pole system consists of a pole attached to a cart that moves over the frictionless track\cite{6313077}. The state space dimensionality is four. The components of state space include, Cart Position (-6,6), Cart-Velocity (-6,6), Pole-Angle(-$\pi$rad, $\pi$rad) and  Pole Angular Velocity(-3,3) 

% In Fig 4. similar to the pendulum environment, we can see images in three different colors. The images in black are the real images. The images in red and blue are images predicted from latent space of trained models. 

% The images in red are regenerated from controllability constrained latent space and images in blue are regenerated from non Controllability Constrained latent space. Controllability Constrained  models give more precise prediction with respect to real time images.
\begin{figure}[t!]
\begin{center}
    \begin{tabular}{p{2cm}p{2cm}p{2cm}}
         \centering Original \\ Dynamics & \centering Constrained \\ Dynamics & \centering Unconstrained \\ Dynamics 
    \end{tabular}
\end{center}
  $t_0$\;\begin{minipage}{\linewidth}
    \frame{\includegraphics[width=0.3\linewidth]{figures/cartpole/original/t1.png}}
    \frame{\includegraphics[width=0.3\linewidth]{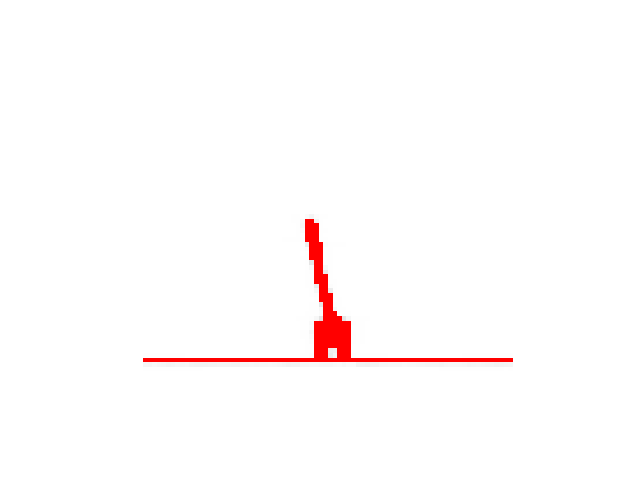}}
    \frame{\includegraphics[width=0.3\linewidth]{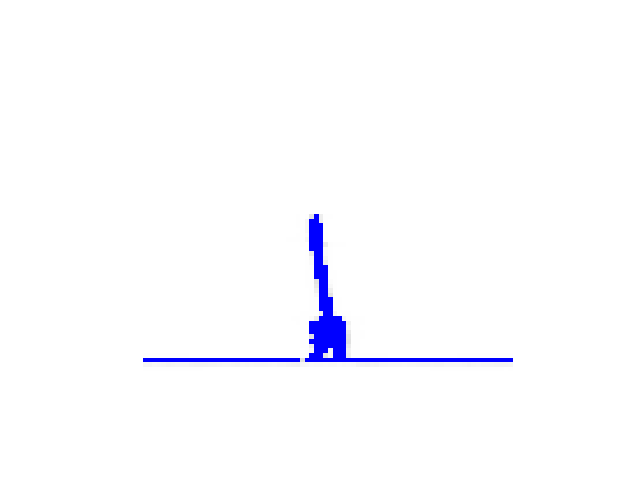}}
  \end{minipage} \centering
 \vspace{0.01cm}\\ 
  $t_2$\;\begin{minipage}{\linewidth}
    \frame{\includegraphics[width=0.3\linewidth]{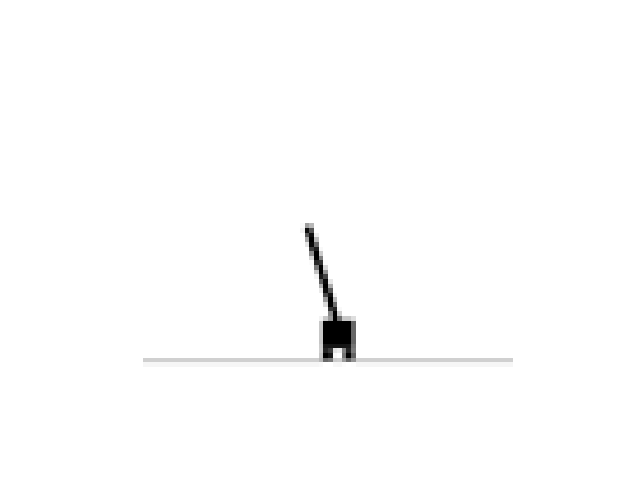}}
    \frame{\includegraphics[width=0.3\linewidth]{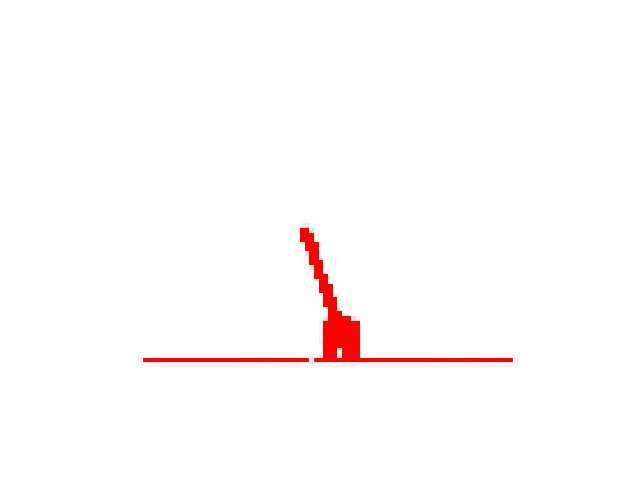}}
    \frame{\includegraphics[width=0.3\linewidth]{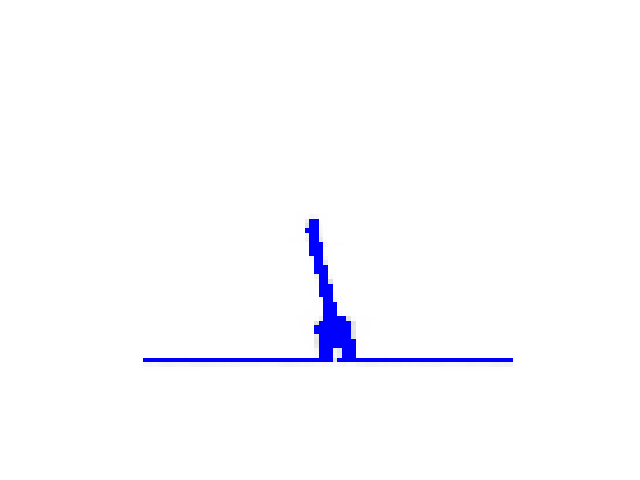}}
  \end{minipage} \vspace{0.01cm}\\
  $t_7$\;\begin{minipage}{\linewidth}
    \frame{\includegraphics[width=0.3\linewidth]{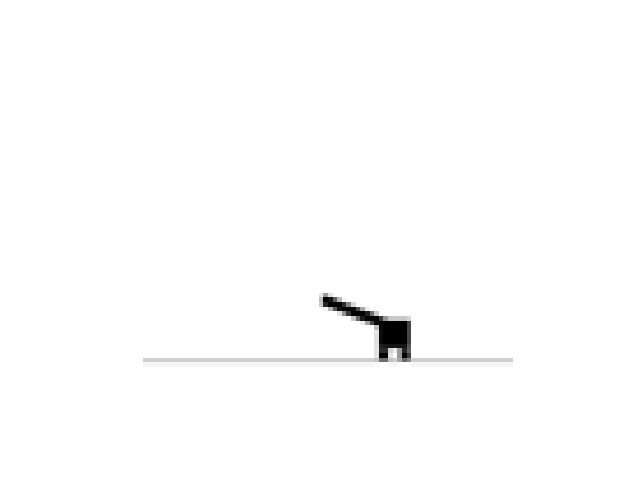}}
    \frame{\includegraphics[width=0.3\linewidth]{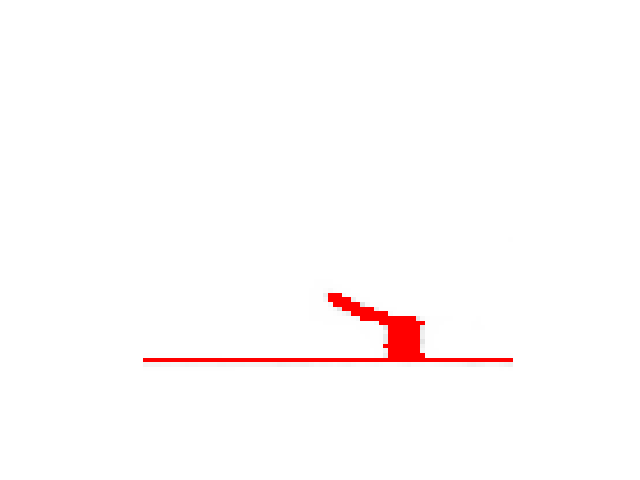}}
    \frame{\includegraphics[width=0.3\linewidth]{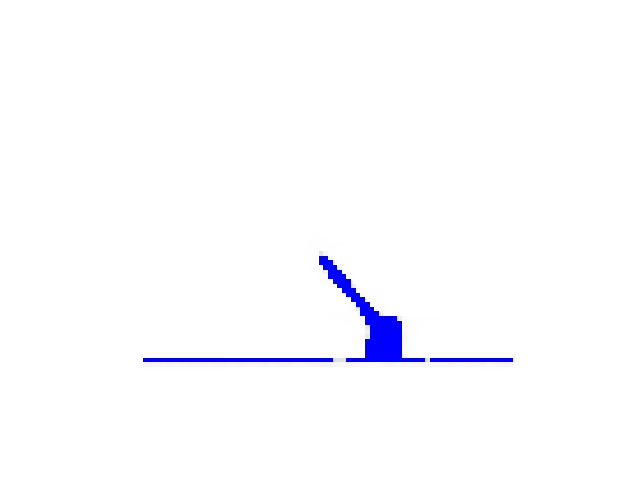}}
  \end{minipage} \vspace{0.01cm}\\
  $t_8$\;\begin{minipage}{\linewidth}
    \frame{\includegraphics[width=0.3\linewidth]{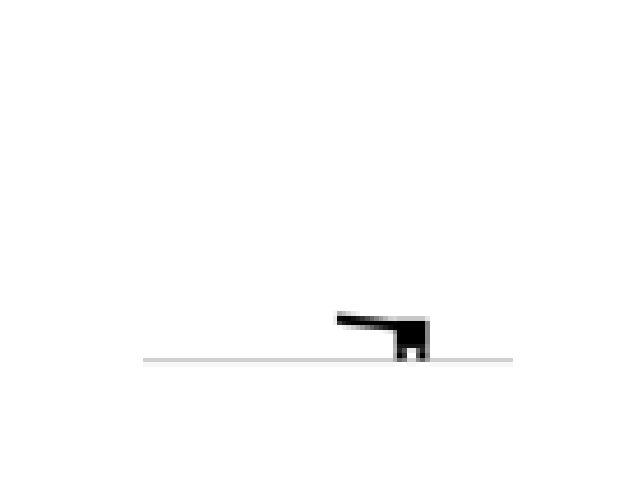}}
    \frame{\includegraphics[width=0.3\linewidth]{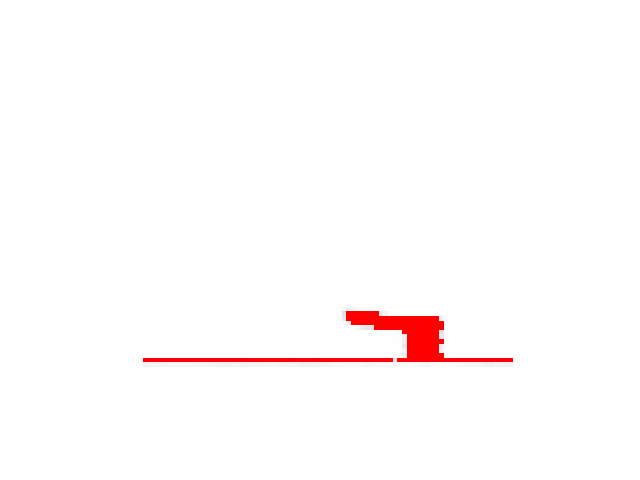}}
    \frame{\includegraphics[width=0.3\linewidth]{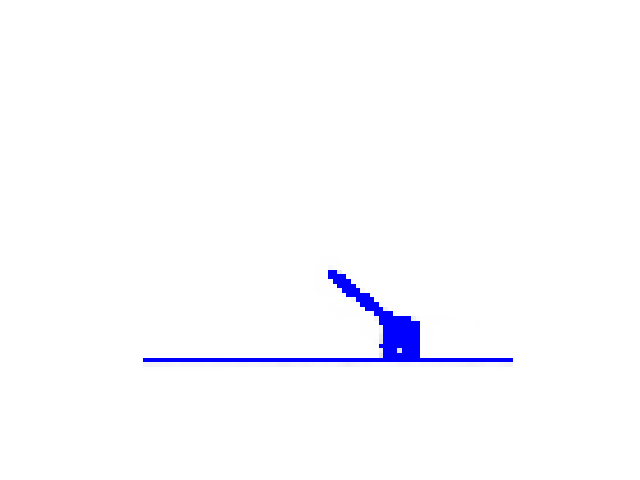}}
  \end{minipage} \vspace{0.01cm}\\
  $t_9$\;\begin{minipage}{\linewidth}
    \frame{\includegraphics[width=0.3\linewidth]{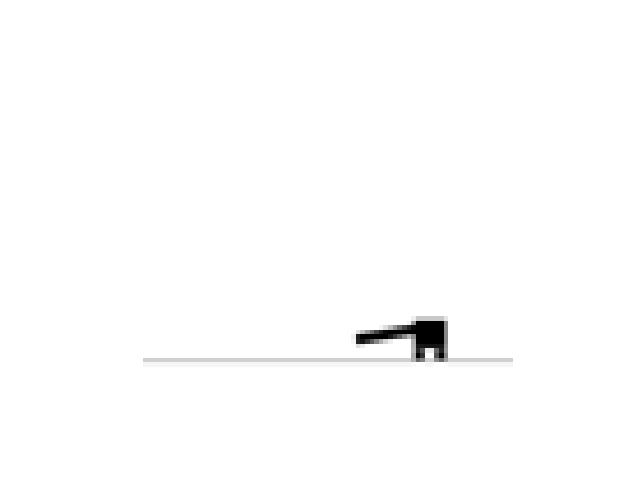}}
    \frame{\includegraphics[width=0.3\linewidth]{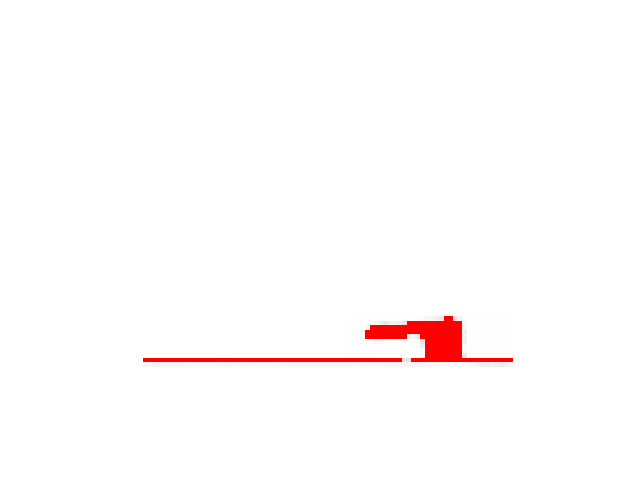}}
    \frame{\includegraphics[width=0.3\linewidth]{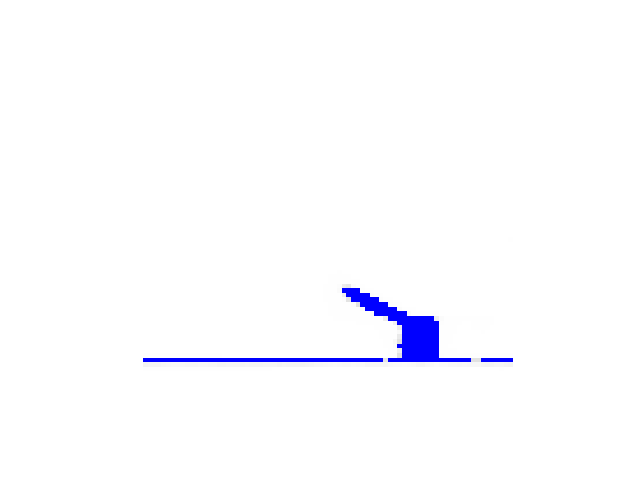}}
  \end{minipage}  \\
  \caption{\label{fig:CartPoleImages} 
  Qualitative comparison in Cart Pole. The column and row meaning is the same as in Figure \eqref{fig:pendImages}.
  The pole starts at $\theta(t_0)=\frac{\pi}{10}\hspace{0.05cm}{\si{\radian}}$ from the top and $\dot{\theta}(t_0)=0\hspace{0.05cm}{\si{\radian\per\second}}$. In this more complicated environment, there is a significant deviation of the baseline solution (blue) from the original dynamics (black) after 7 times steps, while the proposed method (red) closely follows the original dynamics in further steps as well. } 
\end{figure}
\vspace{0.5cm}
% The control task involves balancing the pole in the upright position and making sure it does not fall beyond the range of (-$\pi$/4rad, $\pi$/4rad)
% The real planning cost is concerned, We see a curve similar to the pendulum environment. The best range of $\beta$ values to generate controllability-constrained latent space varies between (0.002, 0.007) with minima close to 0.005. 
% \newpage
\vspace{-0.5cm}
\section{Conclusion and Future Work}\label{sec:conclusion}
In this work, we conceptualize the idea of the enhancement of data-driven models with the controllability property. The implementation of this idea boils down to adding the controllability constraint to the standard VAE objective for learning a dynamics model from data. The resulting controllability-enhanced VAE objective has the same number of parameters to optimize as in the original VAE. That is because the constraint, (the degree of controllability, expressed via the controllability Grammian), is given with the same parameters as in the original unconstrained model. We discovered the controllability-prediction tradeoff, which is regulated by a scalar parameter $\beta$. This proof-of-the-concept study shows that data-driven models may have a small prediction error, which is not enough to guarantee the controllability of the model. The main observation and conclusion are that there exists a positive value of $\beta$, when the performance of a derived controller with a learned model is superior to that for $\beta=0$, as shown in Figure \eqref{fig:pendMPC}. We defer the optimization of the trade-off parameter $\beta$ to future work, where the optimal value of $\beta^*$ will be explored with the dual optimization techniques. A full-fledged method for the derivation of data-driven dynamical models with control-theoretic guarantees for controllability will allow to design of efficient model-based controllers for various critical mission applications such as, neuro-stimulation feedback, robotic surgery, etc.

%%%%%%%%%%%%%%%%%%%%%%%%%%%%%%%%%%%%%%%%%%%%%%%%%%%%%%%%%%%%%%%%%%%%%%%%%%%%%%%%

%%%%%%%%%%%%%%%%%%%%%%%%%%%%%%%%%%%%%%%%%%%%%%%%%%%%%%%%%%%%%%%%%%%%%%%%%%%%%%%%

%%%%%%%%%%%%%%%%%%%%%%%%%%%%%%%%%%%%%%%%%%%%%%%%%%%%%%%%%%%%%%%%%%%%%%%%%%%%%%%%

% \newpage
\section*{ACKNOWLEDGMENT}

% The preferred spelling of the word ÒacknowledgmentÓ in America is without an ÒeÓ after the ÒgÓ. Avoid the stilted expression, ÒOne of us (R. B. G.) thanks . . .Ó  Instead, try ÒR. B. G. thanksÓ. Put sponsor acknowledgments in the unnumbered footnote on the first page.

%%%%%%%%%%%%%%%%%%%%%%%%%%%%%%%%%%%%%%%%%%%%%%%%%%%%%%%%%%%%%%%%%%%%%%%%%%%%%%%%
% {\color{red} We should acknowledge the CoE Small faculty grant. }
The authors thank the graduate students at the CI$^2$ Lab, CoE, SJSU, Tristan Shah and Himaja Papala for proofreading the manuscript. 

\printbibliography

\section*{APPENDIX}\label{sec1:firstappendix}

\subsection{Network Architectures and Hyperparameters }

\subsubsection{Pendulum}
Input: Two $48 \times 48$ images, 15000 training samples of the form $(o_t,a_t,o_{t+1})$,
Action Space: 1-dimensional,
Latent Space: 2-dimensional,
Encoder: 4 layers: Convolution: $16 \times 3 \times 3$; stride (2,2) - Convolution: $32 \times 3 \times 3$; stride (2, 2)  - 256 units - 4 units (2 for mean and 2 for variance),
Decoder: 5 layers: 256 units - 4608 units -  Convolution Transpose: $16 \times 3 \times 3$; stride (2, 2)  - Convolution Transpose: $2 \times 3 \times 3$; stride (2, 2) - Sigmoid Layer,
Transition Dynamics: 3 layers: 100 units - 100 units - 8 units,
MPC Prediction Horizon (H): 10, Control Horizon: 10,
Number of Models (q): 20,
Learning Rate: 0.0003,
Optimizer: 'ADAM' \cite{kingma2014adam},
Cost Matrices: $Q=I_{n\times n}$, $R= 0.01$.

\subsubsection{CartPole}
The architecture is inspired by. \cite{levine2019prediction}
Input: Two $80 \times 80$ images. 15000 training samples of the form $(o_t,a_t,o_{t+1})$,
Action Space: 1-dimensional,
Latent Space: 4-dimensional,
Encoder: 6 layers: Convolution: $32 \times 5 \times 5$; stride (1,1) - Convolution: $32 \times 5 \times 5$; stride (2, 2)-Convolution: $32 \times 5 \times 5$; stride (2, 2)-Convolution: $10 \times 5 \times 5$; stride (2, 2)  - 200 units - 8 units (4 for mean and 4 for variance),
Decoder: 7 Layers: 200 units - 1000 units -  Convolution Transpose: $32 \times 5 \times 5$; stride (1, 1)  - Convolution Transpose: $32 \times 5 \times 5$; stride (1, 1) - Convolution Transpose: $32 \times 5 \times 5$; stride (1, 1) - Convolution Transpose: $2 \times 5 \times 5$; stride (1, 1)-Sigmoid Layer,
Transition Dynamics: 3 layers: 100 units - 100 units - 24 units,
MPC Prediction Horizon: 10, Control Horizon: 10.
Number of Models (q): 8,
Learning Rate: 0.0001,
Optimizer: 'ADAM' \cite{kingma2014adam},
Cost Matrices : $Q=I_{n\times n}$, $R=1$.

\subsection{Additional Training Details}
% The Neural Network model has three modules, the encoder, decoder, and transition network. The encoder is used to extract out the latent representation of $o_{t}$  and $o_{t+1}$ as $z_{t}$ and   $z_{t+1}$. Then transition network maps $z_{t}$ and $a_{t}$ to $\hat{z}_{t+1}$ similar to Eqn \ref{eq:latent dynamics}. This task aims to minimize $L_{e2c}$ and maximize the minimal SVD value of the Grammian to generate a more constrained state space  for optimal control. 

% The main challenge was to find the point of convergence for all the $\beta$ values.Due to which results were randomized. To solve this problem, all the models, including  the Standard as well as Controllability Constrained models, were trained over the same base model \cite{zhuang2020comprehensive}. The above step was repeated 'k' times with 'k' different base models.

% {\color{red} in the main text we provide the introduction, the problem formulation, the method, and the experiments. please move all the technical details, number of repetitions etc to appendix.}
% The challenging task in this step was to find a point of convergence for every training round  utilizing a particular $\beta$ value. To solve the above problem both the Standard as well Controllability Constrained models were trained over the  same base model. The process was repeated k times for every environment with 'k' base models \cite{zhuang2020comprehensive}. 
To generate Fig \ref{fig:pendMPC} for every $\beta$ value 'q' sets were created. To compute the final original planning cost for the baseline as well as controllability-constrained models, an average of all the "original total cost" values for every (q,$\beta$) model was taken.

\end{document}